\documentclass[12pt,twoside]{amsart}
\usepackage{amscd,amssymb,righttag}
\def\R{{\mathbb R}}

\def\N{{\mathbb N}}
\def\s{{\mathbb S}}
\def\e{{\epsilon}}

\newtheorem{thm}{Theorem}[section]
\newtheorem{corol}[thm]{Corollary}

\newtheorem{lemma}[thm]{Lemma}
\newtheorem{prop}[thm]{Proposition}
\theoremstyle{definition}
\newtheorem{defin}[thm]{Definition}

\theoremstyle{remark}
\newtheorem{remark}[thm]{Remark}

\newtheorem{example}[thm]{Example}
\newtheorem{examples}[thm]{Examples} 
\numberwithin{equation}{section}
\begin{document}

\begin{center}
\renewcommand{\thefootnote}{\fnsymbol{footnote}}

{\Large\bf On the geometry of similarity search:
dimensionality curse and concentration of measure\footnote[3]{Partially 
supported by the
Marsden Fund grant VUW703 of the Royal Society
of New Zealand.}}
\\[.6cm]

{\large Vladimir Pestov\footnote[1]
{E-mail: vova@@mcs.vuw.ac.nz \\ \indent
 URL: http://www.vuw.ac.nz/$^\sim$vova}} 
\\[.5cm]

\footnotetext{Preprint data: RP-99-01, School of Mathematical and Computing
Sciences, Victoria University of Wellinton, January 1999.}

{School of Mathematical and Computing Sciences, 
Victoria University of
Wellington, \\
 P.O. Box 600,
Wellington, New Zealand.}
\\[1cm]

\end{center}
\hrule
\vskip .4cm

{\footnotesize\bf Abstract}

\begin{quote}
{\footnotesize
\noindent
We suggest that the curse of dimensionality affecting the similarity-based
search in large datasets is a manifestation of the
phenomenon of concentration of measure on high-dimensional structures.
We prove that, under certain geometric assumptions on the query domain
$\Omega$ and the dataset $X$, if $\Omega$ satisfies the so-called
concentration property, then
for most query points $x^\ast$ the ball of radius $(1+\e)d_X(x^\ast)$
centred at $x^\ast$ contains either all points of $X$ or else
at least $C_1\exp(-C_2\e^2n)$ of them.
Here $d_X(x^\ast)$ is the distance from $x^\ast$ 
to the nearest neighbour in $X$ and
$n$ is the dimension of $\Omega$.}
\end{quote}
\smallskip

{\footnotesize\bf Keywords}

\begin{quote} {\footnotesize
Data structures, databases, information retrieval,
computational geometry, performance evaluation}
\end{quote}
\vskip .4cm

\hrule 

\vskip 1cm

\section{Introduction}
As the size of datasets in existence grows
at an amazing rate (see e.g.
Section 4.1 in \cite{SSU}) and workloads become ever more sophisticated, 
algorithms for similarity-based data retrieval often
slow down exponentially with dimension, sometimes
reducing to an exhaustive search 
(`the curse of dimensionality') \cite{BGRS,BWY,pyr2,WSB}. 
It is important to try and understand the common geometric nature
of the dimensionality curse for a great variety of different, often
non-euclidean, metric spaces representing data structures
\cite{Brin,CPRZ,CPZ,U}.

In this Letter we suggest that the curse of dimensionality
is a manifestation of the phenomenon of concentration of measure on
high-dimensional structures. 

This phenomenon is an important
discovery of modern analysis, observed in a wide range of
situations \cite{GrM,M,MS,Ta}. 
Roughly speaking, a set $\Omega$ equipped with a distance and a probability 
measure has the concentration property if already for small values
of $\e>0$ the `$\e$-fattening' of every subset 
containing at least 1/2 of all elements of $\Omega$ contains
all points of $\Omega$ apart from a set of almost vanishing measure 
$\alpha(\e)$. Here $\alpha$
is the so-called concentration function of $\Omega$.
Many `naturally occuring'
high-dimensional structures possess the concentration property:
the $n$-dimensional sphere
$\s^n$, Euclidean unit ball ${\mathbb B}^n$, Hamming cube $\{0,1\}^n$, 
groups of permutations
$S_n$ all have concentration functions of the form 
$\alpha(\e)=O(1)\exp(-O(1)\e^2n)$.

Here we will address just one aspect of 
`dimensionality curse,' informally described
in \cite{BWY} as follows: 
\begin{quote}
\small
`It seems ... that this {\rm [exponential]}
complexity might be inherent in any algorithm for solving closest point
problems because a point in a high-dimensional space can have many
``close'' neighbours.'
\end{quote}
To formalise this account, we borrow
a concept from \cite{BGRS}.
A similarity query is called $\e$-unstable for an $\e>0$ if most
points if the dataset $X$ are at a distance $<(1+\e)d_X(x^\ast)$
from the query point $x^\ast$, where $d_X$ denotes the distance
to the nearest neighbour in the dataset $X$. 
Query instability was shown in \cite{BGRS} to occur under some
probability assumptions on the query distribution, and it was
argued that asking unstable queries is partly responsible for the
dimensionality curse. 
It seems to us that even more important is
a `local' version of query instability, where
the number of data points located at a distance $<(1+\e)d_X(x^\ast)$ 
from a query point $x^\ast$ grows 
exponentially in the dimension of
the query domain. If such an effect prevails in a given workload,
then answering the range query of radius $(1+\e)d_X(x^\ast)$
obviously takes an average expected exponential time, even though the
query may be globally stable. 

In our model, a dataset $X$ is a finite metric
subspace of a metric space $\Omega$ of query points, the latter being
equipped with a probability measure reflecting the query distribution. 
We assume that $\Omega$ has the concentration property in the sense
that $\alpha(\e)=O(1)\exp(-O(1)\e^2n)$, where $n$ is the 
`dimension' of the query domain $\Omega$.
Our assumption on the way $X$ sits in $\Omega$
is of a homogeneity type: the 
radii of open balls centred at $x$ and having measure $1/2$ are
(almost) the same for all datapoints $x$.

Under such assumptions we prove that if $\e>0$, then for all query
points $x^\ast\in\Omega$, apart from a set of measure $O(1)\exp(-O(1)\e^2n)$,
the open ball of radius $(1+\e)d_X(x^\ast)$ centred at $x^\ast$
contains either all points of the dataset 
$X$ or else at least
$C_1\exp(C_2\e^2n)$ of them for some $C_1,C_2>0$.
Thus, a typical range query of radius $(1+\e)d_X(x^\ast)$ is
either unstable or takes an exponential
time to answer. In particular, most queries are unstable if the size
of $X$ grows subexponentially in $n$.

In Conclusion we explain a possible constructive significance of our results.

\section{Similarity workloads}
Our model builds on the approaches of 
\cite{CPRZ,CPZ} and \cite{HKP}.
A {\it similarity workload} is a quadruple $(\Omega, d,\mu, X)$,
where
\begin{enumerate}
\item $\Omega$ is a (possibly infinite) set called the {\it domain},
whose elements are {\it query points.}
\item $ d$ is a metric on $\Omega$, the {\it dissimilarity measure.}
\item $\mu$ is a Borel probability measure on the metric space
$(\Omega, d)$, reflecting the
query point distribution.
\item $X$ is a finite subset of $\Omega$, called the {\it instance,} or
the {\it dataset} proper, whose elements are {\it data points}.
\end{enumerate} 

Recall that a triple $(\Omega, d,\mu)$ formed by a metric space
$(\Omega, d)$ and a probability Borel measure $\mu$ on it is called a
{\it probability metric space.} Thus, a similarity workload 
is a probability metric 
space $\Omega$ together with a distinguished finite
metric subspace $X$. 

{\it Similarity queries} are of two major types: a 
{\it range query} centred at $x^\ast\in\Omega$ of radius
$\e>0$ (the set of all $x\in X$ with $ d(x,x^\ast)<\e$), and
a {\it $k$-nearest neighbours \textrm{(}$k$-NN\textrm{) }
query} centred at $x^\ast\in\Omega$, where $k\in\N$.

Following \cite{BGRS}, we
say that a similarity query centred at $x^\ast$ is $\e${\it -unstable}
for an $\e>0$ if 
\[ \left\vert\{x\in X\colon  d(x^\ast,x)\leq
(1+\e)d_X(x^\ast) \}\right\vert > \frac {\vert X\vert} 2,\]
where $d_X(x^\ast)=\min\{d(x^\ast,x)\colon x\in X\}$ is the distance from 
$x^\ast$ to the nearest neighbour in $X$.
In \cite{BGRS} the following new type
of queries is proposed.

\begin{itemize}
\item An {\it $\e$-radius nearest neighbours query} centred at a
point $x^\ast\in\Omega$, where $\e>0$, is a range query centred
at $x^\ast$ of the radius $(1+\e)d_X(x^\ast)$.
\end{itemize}

\section{The concentration phenomenon}

The {\it concentration function,}
$\alpha=\alpha_\Omega$, of a probability metric space $\Omega$ is defined 
for each $\e>0$ by
\begin{equation} 
\alpha_\Omega(\e)=1-\inf\left\{\mu\left({\mathcal O}_\e(A)\right)
\colon A\subseteq \Omega \mbox{ is Borel and }
\mu(A)\geq \frac 1 2\right\}
\end{equation}
and $\alpha_\Omega(0)=1/2$. It is a decreasing function in $\e$.

A family $(\Omega_n)_{n=1}^\infty$ of probability metric spaces is called
a {\it L\'evy family} if for each $\e>0$, $\alpha_{\Omega_n}(\e)\to 0$ as
$n\to\infty$,
and a {\it normal L\'evy family} (with constants $C_1,C_2>0$) if for
all $n$ and $\e>0$
\[\alpha_{\Omega_n}(\e)\leq C_1e^{-C_2\e^2n}.\]
All the families listed below are normal L\'evy families, 
see \cite{GrM,M,MS} for exact values of constants and further
examples.

\begin{examples}
\label{ex}
(1) The $n$-dimensional unit spheres ${\mathbb S}^n$ equipped
with the (unique) rotation-invariant probability measure and the
geodesic distance. (2) The same, with the Euclidean distance.
(3) The Hamming cubes $\{0,1\}^n$ of all binary strings
of length $n$, equipped with the normalised
Hamming distance 
$d(s,t)= \frac 1 n \vert \{i\colon s_i\neq t_i\}\vert$ and the
normalised counting measure
$\mu_\sharp(A)=\vert A\vert/\vert X\vert$. 
(4) The groups $SO(n)$ of $n\times n$ 
orthogonal matrices with determinant $1$, 
equipped with the geodesic distance
and the Haar measure. (5) The Euclidean balls ${\mathbb B}^n$
with the $n$-volume and Euclidean distance. (6) The tori ${\mathbb T}^n$
with the normalised geodesic distance and product measure.
(7) The hypercubes $[0,1]^n$ with the normalised Euclidean (or $l_1$)
distance. 
\end{examples}

More L\'evy families can be obtained using operations described in
\cite{GrM}, Sect. 2.

Let $f\colon \Omega\to \R$ be a Lipschitz-1 function:
\[\forall x,y\in\Omega,~~\vert f(x)-f(y)\vert\leq d(x,y). \]
Denote by $M$ a median (or L\'evy mean) of $f$, that is,
a real number with 
\[\mu(\{x\in X\colon f(x)\leq M\})
=\mu(\{x\in X\colon f(x)\geq M\}).\]
\begin{prop}
For every $\e>0$,
$\mu\left(f^{-1}(M-\e,M+\e) \right)\geq 1-2\alpha(\e)$.
\label{lip}
\qed
\end{prop}
The {\it phenomenon of concentration of measure on high-dimensional
structures} refers to the above situation, in which the function $f$
`concentrates near one value.'

See \cite{GrM,M,MS,Ta}.

\section{Concentration and similarity workloads}

Let $\Omega$ be a probability metric space with the
concentration function $\alpha=\alpha_{\Omega}$.
The following is quite immediate.

\begin{lemma} Let $A\subseteq\Omega$, $\delta>0$, and
$\mu(A)>\alpha(\delta)$.
Then $\mu({\mathcal O}_\delta(A))> 1/2$.
\label{half}
\qed
\end{lemma}

\begin{lemma} Let $\delta>0$, and let
$\gamma$ be a collection of 
subsets $A\subseteq\Omega$ of measure $\mu(A)\leq\alpha(\delta)$ each,
satisfying $\mu(\cup\gamma)\geq 1/2$. 
Then the $2\delta$-neighbourhood of every point 
$x\in \Omega$, apart from a set of measure at most
$\frac 12\alpha(\delta)^{\frac 12}$,
 meets at least $\lceil \frac 1 2 \alpha(\delta)^{-\frac 12}\rceil$
 elements of $\gamma$.
\label{tech}
\end{lemma}

\begin{proof} 
Partition
$\gamma$ into a collection of pairwise
disjoint subfamilies $\gamma_i$, $i\in I$ in such a way that for
every $i$, $\alpha(\delta)\leq\mu(A_i)<
2\alpha(\delta)$, where $A_i= \cup\gamma_i$. Clearly, 
$(1/4)\alpha(\delta)^{-1}\leq \vert I\vert \leq (1/2)\alpha(\delta)^{-1}$. 
Select a subset $J\subseteq I$ with 
$\vert J\vert=\lceil \frac 12\alpha(\delta)^{-\frac 12} \rceil$.
Lemma \ref{half} implies that 
\[\mu\left({\mathcal O}_{2\delta}(A_i)\right)\geq 
\mu\left({\mathcal O}_{\delta}\left({\mathcal O}_{\delta}A_i\right)\right) 
\geq 1-\alpha(\delta),\]
and therefore $\cap_{i\in J}({\mathcal O}_{2\delta}(A_i)$ has measure at
most $1-\vert J\vert \alpha(\delta)$.
\end{proof}

Let $(\Omega,d,\mu,X)$ be a similarity workload, with $\alpha$  
as above. Denote by $M$ a median value of the function
$d_X$ (distance to $X$) on $\Omega$.

\begin{lemma}
\label{sim} Let $\delta>0$.
Then for all points $x^\ast\in\Omega$, except for a set of total mass at
most $2\alpha(\delta)$, the distance to the nearest 
neighbour in $X$ is in the interval $(M-\delta,M+\delta)$.
\end{lemma}

\begin{proof} The function $x^\ast\to d_X(x^\ast)$ is 
Lipschitz-1 on $\Omega$, and Prop. \ref{lip} applies. 
\end{proof}

\begin{defin}
Let $(\Omega, d,\mu,X)$ be a similarity workload. For an $x\in X$,
denote by $R_x$ the maximal radius of an open ball in $\Omega$ 
centred at $x$ of measure $\leq 1/2$.
Let $\e>0$.
We say that $X$ is {\it weakly $\e$-homogeneous} in $\Omega$ if all radii
$R_x$, $x\in X$ belong to an interval of length $<\e$. 
\end{defin}

\begin{examples} (1) $X$ is weakly $\e$-homogeneous for 
every $\e>0$ if the group of motions
preserving the measure acts transitively on $\Omega$. Such are spaces
1-4, 6 in Example \ref{ex}.
(2) A subspace $X$ of the ball ${\mathbb B}^n$ is weakly
$\e$-homogeneous if $X$ is contained in a spherical shell of thickness
$\e$.
(3) If we independently throw in $\Omega$ $N$ points
$x_1,x_2,\dots,x_N$, distributed
with respect to the measure $\mu$, then one can show
that, with probability
$\geq 1-2N\alpha(\e/2)$, the dataset $X=\{x_1,\dots,x_N\}$ is 
weakly $\e$-homogeneous.
\end{examples}

\section{Query instability: local and global}
\begin{thm}
\label{main}
Let $(\Omega, d,\mu,X)$ be a similarity workload. Denote by
$M$ a median value of the distance from a query point in
$\Omega$ to its nearest neighbour in $X$. Let $0<\e<1$,
and assume that the instance $X$ is weakly 
$(M\e/6)$-homogeneous in $\Omega$.

Then for all points $x^\ast\in\Omega$, apart from a set of total mass
at most $3\alpha(M\e/6)$,
the open ball of radius $(1+\e)d_X(x^\ast)$ centred at $x^\ast$
contains at least
\begin{equation}
\min\left\{\vert X\vert,~
\lceil\frac 1 {2\alpha(M\e/6)^{\frac 12}}\rceil\right\}
\label{N}
\end{equation}
elements of $X$.
\end{thm}

\begin{proof} 
Denote by $R$ the minimum of the radii
$R_x, x\in X$. Let $\Delta=R-M$.

(1) If $\Delta>M\e/6$, then by Lemma \ref{half}
the measure of the ball ${\mathcal O}_M(x)$ cannot exceed 
$\alpha(\Delta)\leq\alpha(M\e/6)$, for otherwise
the measure of ${\mathcal O}_R(x)$ would be $>1/2$.
In particular, $\vert X\vert\geq \frac 12\alpha(M\e/6)^{-1}$.
According to Lemma \ref{tech} applied to the
balls ${\mathcal O}_M(x)$, $x\in X$ with $\delta=M\e/6$,
for all $x^\ast\in\Omega$ apart from
a set of measure $\leq \frac 1 2\alpha(M\e/6)^{\frac 12}$, the
$(M\e/3)$-neighbourhood of $x^\ast$ meets at least 
$\lceil \frac 1 2\alpha(M\e/6)^{-\frac 12}\rceil$ of such balls.

(2) If $\Delta\leq M\e/6$ (in particular, if 
$\vert X\vert< \frac 12\alpha(M\e/6)^{-\frac 12}\leq 
\frac 12\alpha(M\e/6)^{-1}$, cf. the previous paragraph), then
$R_x+M\e/6\leq M(1+\e/2)$.
Denote by $X'$ a subset of $X$ of cardinality
$\min\{\vert X\vert, \frac 12\alpha(M\e/6)^{-\frac 12} \}$.
Since the measure of every ball ${\mathcal O}_{M(1+\e/2)}(x)$ is
at least $1-\alpha(M\e/6)$, for all
$x^\ast\in\Omega$ apart from a set of measure
$\leq \frac 12\alpha(M\e/6)^{\frac 12}$,
the $(M\e/2)$-neighbourhood of $x^\ast$ meets every ball
${\mathcal O}_M(x)$, $x\in X'$. 

As a consequence of Lemma \ref{sim} with $\delta=M\e/4$,
for all $x^\ast\in\Omega$ apart
from a set of measure at most $2\alpha(M\e/4)$, one has
$\vert d_X(x^\ast)-M\vert<M\e/4$ and therefore
$M(1+\e/2)\leq d_X(x^\ast)(1+\e)$.
It remains to notice that  
$\frac 12\alpha(M\e/6)^{\frac 12}+2\alpha(M\e/4)\leq
3\alpha(M\e/6)^{\frac 12}$.
\end{proof}

\subsection*{Asymptotic results}
Let $(\Omega_n, d_n,\mu_n,X_n)$ be an infinite collection of
workloads. Denote by $M_n$ the median distances  
from points of $\Omega_n$ to their nearest neighbours in $X_n$.
We make the following standing assumptions.
\begin{enumerate}
\item[(1)] The query domains $(\Omega_n, d_n,\mu_n)$ form a 
{\it normal L\'evy family.}
\item[(2)] The values $M_n$ are bounded away from zero:
$M_n\geq M>0$ for all $n\in\N$.
\end{enumerate}

\begin{remark} The latter condition is only violated in very densely
populated domains. For example, if
$\Omega_n=\s^n$, then (2) is satisfied whenever the size of $X_n$
is not superexponential in $n$. For $\Omega_n$ finite 
(2) is satisfied if 
$\vert X_n\vert\leq \alpha_{\Omega_n}(M)\cdot\vert\Omega_n\vert$.
\end{remark}

Now let $0<\e<1$. 

\begin{enumerate} 
\item[(3)]
All the instances $X_n$ are
weakly $(M_n\e/6)$-homogeneous in $\Omega_n$.
\end{enumerate}

\begin{corol}  Under the assumptions {\rm (1)-(3),}
for all query points $x^\ast\in\Omega_n$, apart from a set of
measure 
$O(1)\exp(-O(1)M^2\e^2n)$,
the open ball of radius
$(1+\e)d_X(x^\ast)$ centred at $x^\ast$ contains
either all elements of $X$ or else at least
$C_1\exp(C_2M^2\e^2n)$ of them for some constants $C_1,C_2>0$ depending
only on the family $(\Omega_n)_{n=1}^\infty$.
\label{local}
\qed
\end{corol}

\begin{corol} Under the assumptions {\rm (1)-(3),} for all query
points $x^\ast$, apart from a set of measure 
$O(1) \exp(-O(1)M^2\e^2n)$, the
$\e$-radius nearest neighbours query centred at 
$x^\ast$ either is unstable or takes
an exponential time (in $n$) to answer.
\qed
\end{corol}

\begin{corol} In addition to {\rm (1)-(3),} let the size of
$X_n$ grow subexponentially in $n$.
Then for all query points $x^\ast\in\Omega_n$, apart 
from a set of measure
$O(1)\exp(-O(1)M^2\e^2n)$,
 the similarity query centred at $x^\ast$ is
$\e$-unstable: all
points of $X_n$ are at a distance $< (1+\e)d_X(x^\ast)$ from $x^\ast$. 
\qed\end{corol}

\begin{example} It is easy to construct sequences of workloads in which
most of similarity queries are $1$-stable and yet for every
$\e>0$ most of the $\e$-radius NN queries take time
$C_1\exp(C_2\e^2n)$ to answer.

Let $\delta>0$ be arbitrary.
In a probability metric space $\Omega$
choose a maximal subset $X$ with the property that every
two different elements of $X$ are at a distance $>\delta$ from each other.
It is easy to see 
that centres of all $1$-unstable similarity queries in the workload
$(\Omega,X)$ are contained in some
ball of radius $4\delta$. 
Applying this procedure to every member of a normal L\'evy family
of homogeneous spaces of constant diameter $D$ 
(a typical situation) and choosing
$\delta<D/8$, we obtain a desired sequence of workloads, because one
can then prove that $\liminf M_n\geq \delta/2$.
\end{example}

\section{Conclusion} Our model links the 
`curse of dimensionality' in multidimensional 
datasets to the phenomenon of 
concentration of measure on high-dimensional structures.
All our assumptions on the query domain
$\Omega$ and the dataset $X$ are purely geometric.
Our estimates are by no means optimal, as we just aimed at
deriving exponential lower bounds in a wide variety of situations.
We believe that the most general case (absence of homogeneity in
any form) can be included in the picture as well and 
will address the issue in the
future work. Other important directions for research are to apply
the concentration phenomenon to indexability theory \cite{HKP}
and to performance analysis of concrete hierarchical tree index structures
\cite{BWY,Brin,CPRZ,CPZ,U}.

A possible constructive significance of our results is as follows. In 
practice, geometrically optimal dissimilarity measures are being 
routinely replaced with less precise distances that are computationally 
cheaper, with a view of subsequently discarding false hits. 
Such distances would in general lead to sharper 
concentration effects on the same measure space.
It is therefore conceivable that using computationally more expensive 
distances will result in an overall speed-up.

\section*{Acknowledgements}
I am grateful to Paolo Ciaccia for introducing me
to the problematics of similarity-based information storage and retrieval,
as well as for his hospitality and stimulating discussions
during my visit to the University of Bologna in June 1998.


\begin{thebibliography}{100}

\bibitem{pyr2} S. Berchtold, C. B\"ohm, D.A. Keim, and H.-P. Kriegl,
{\it A cost model for nearest neighbour search in
high-dimensional data space,} PODS'97 (Tucson, AZ), 78--86.

\bibitem{BGRS}
K. Beyer, J. Goldstein, R. Ramakrishnan, and U. Shaft,
{\it When is ``nearest neighbor'' meaningful?,}
Technical paper no. 226, CS dept., Univ. Wisconsin-Madison,
to appear in: ICDT-99.

\bibitem{BWY} J.L. Bentley, B.W. Weide, and A.C. Yao,
{\it Optimal expected-time algorithms for closest point problems,}
ACM Trans. Math. Software {\bf 6} (1980), 563--580.

\bibitem{Brin}
S. Brin, {\it Near neighbor search in large metric spaces,}
in: Proc. of the 21st VLDB International Conf.,  Zurich,
Switzerland, Sept. 1995, pp. 574--584.

\bibitem{CPRZ} 
P.~ Ciaccia, M.~ Patella, F.~ Rabitti, and P.~ Zezula,
{\it Performance of $M$-tree, an access method for similarity
search in metric spaces,} EC ESPRIT report,
24 February 1997, 25 pp., downloadable from
{\tt http://www.ced.tuc.gr/hermes}

\bibitem{CPZ} 
P.~ Ciaccia, M.~ Patella, and P.~ Zezula,
{\it A cost model for similarity queries in metric spaces,} in:
Proc. 17-th Annual ACM Symposium on Principles of Database Systems
(PODS'98), Seattle, WA, June 1998, pp. 59--68.

\bibitem{GrM}
M. Gromov and V.D. Milman,
{\it A topological application of the isoperimetric inequality,}
Amer. J. Math. {\bf 105} (1983), 843--854.

\bibitem{HKP}
J.M. Hellerstein, E. Koutsoupias, and C.H. Papadimitriou,
{\it On the analysis of indexing schemes,} in: PODS'97, Tucson, AZ,
pp. 249--256.

\bibitem{M} V.D. Milman, {\it The heritage of P.L\'evy in
geometric functional analysis,} Ast\'erisque {\bf 157-158}
(1988), 273--301.

\bibitem{MS} V.D. Milman and G. Schechtman,
{\it Asymptotic Theory of Finite Dimensional Normed Spaces,}
Lecture Notes in Math. {\bf 1200}, Springer-Verlag, 1986.

\bibitem{SSU} A. Silberschatz, M. Stonebraker, and
J. Ullman (eds.), {\it Database research: achievements and
opportunities into the 21st century,}  Report of an NSF Workshop on
the Future of Database Systems Research, May 26--27, 1995. 

\bibitem{Ta} M. Talagrand, {\it Concentration of measure and
isoperimetric inequalities in product spaces,} Publ. Math. IHES
{\bf 81} (1995), 73--205.

\bibitem{U} J.K. Uhlmann,
{\it Satisfying general proximity/similarity queries with metric
trees,} Information Processing Lett.  {\bf 40} (1991), 175--179.

\bibitem{WSB} 
R. Weber, H.-J. Schek, and S. Blott,
{\it A quantatitive analysis and performance study for 
similarity-search methods in high-dimensional spaces,}
in: Proceedings of the 24-th VLDB Conference, New York, 1998,
pp. 194--205.

\end{thebibliography}
\end{document}